\documentclass[copyright,creativecommons]{eptcs}
\usepackage{wrapfig}
\usepackage{amsmath,amsthm,yhmath}
\hypersetup{colorlinks}
\usepackage{thm-restate}

\usepackage{microtype}

\usepackage{nico}
\usepackage{qctl}
\usepackage{macrosatlsc}
\usetikzlibrary{calc,shapes}
\usepackage{multirow}

\title{Satisfiability of \ATL with strategy contexts}
\author{Fran\c{c}ois Laroussinie
\institute{LIAFA -- Univ.\ Paris Diderot \& CNRS}
\email{francoisl@liafa.univ-paris-diderot.fr}
\and
Nicolas Markey
\institute{LSV -- ENS Cachan \& CNRS}
\email{markey@lsv.ens-cachan.fr}
}
\def\titlerunning{Satisfiability of \ATL with strategy contexts}
\def\authorrunning{Fran\c{c}ois Laroussinie, Nicolas Markey}
\begin{document}
\bgroup
\let\footnotemark\relax
\thanks{\leftskip=0em\relax \noindent
  This work was partly supported by ERC Starting
  grant EQualIS (308087) and by European project Cassting (FP7-ICT-601148).}
\egroup
\maketitle


\begin{abstract}
Various extensions of the temporal logic \ATL have recently been
introduced to express rich properties of multi-agent systems. Among
these, \ATLsc~extends \ATL with \emph{strategy contexts},
while Strategy Logic has \emph{first-order quantification} over
strategies.
There is a price to pay for the rich expressiveness of these logics:
model-checking is non-elementary, and satisfiability is undecidable.

We prove in this paper that satisfiability is decidable in several
special cases. The most important one is when restricting to
\emph{turn-based} games. We~prove that decidability also holds for concurrent
games if the number of moves available to the agents is
bounded. Finally, we~prove that restricting strategy quantification to
memoryless strategies brings back undecidability.

\end{abstract}

\section{Introduction}
\label{sec-intro}

Temporal logics are a convenient tool to reason about computerised
systems, in particular in the setting of
verification~\cite{Pnu77,CE82,QS82a}. When systems are interactive,
the models usually involve several agents (or~players), and relevant
properties to be checked often question the existence of
\emph{strategies} for these agents to achieve their goals. To~handle
these, \emph{alternating-time temporal logic} was
introduced, and its algorithmic properties were studied:
model checking is \PTIME-complete~\cite{AHK02}, while satisfiability was settled
\EXPTIME-complete~\cite{jlc16(6)-WLWW}. 

While model checking is tractable, \ATL still suffers from a lack of
expressiveness. Over the last five years, several extensions or
variants of \ATL have been developed, among which \emph{\ATL with
  strategy contexts}~\cite{BDLM09} and \emph{Strategy
  Logic}~\cite{CHP07b,MMV10a}. The model-checking problem for these
logics has been proved non-elementary~\cite{DLM10,DLM12}, while
satisfiability is undecidable, both when looking for finite-state or
infinite-state models~\cite{MMV10a,TW12}. Several fragments of
these logics have been defined and studied, with the aim of preserving
a rich expressiveness and at the same time lowering the complexity of
the decision problems~\cite{WHY11,MMPV12,HSW13}. 

In this paper we prove that satisfiability is decidable (though with
non-elementary complexity) for the full logic \ATLsc (and~\SL) in two
important cases: first, when satisfiability is restricted to
turn-based games (this solves a problem left open in~\cite{MMV10a}
for~\SL), and second, when the number of moves available to the
players is bounded. We~also consider a third variation, where
quantification is restricted to \emph{memoryless} strategies; in~that setting,
the satisfiability problem is proven undecidable, even for turn-based
games. 

\looseness=-2
Our results heavily rely on a tight connection between \ATLsc and
\QCTL~\cite{DLM12}, the extension of \CTL with quantification over atomic
propositions. For instance, the \QCTL formula $\exists p.\ \phi$ states that
it~is possible to label the unwinding of~the model under consideration with
proposition~$p$ in such a way that $\phi$~holds. This labeling with additional
proposition allows us to mark the strategies of the agents and the
model-checking problem for \ATLsc can then be reduced to the model-checking
problem for~\QCTL.
However, in this transformation, the resulting \QCTL formula depends
both on the \ATLsc formula to be checked and on the game where the
formula is being checked. This way, the procedure does not extend to
satisfiability, which is actually undecidable. We~prove here that this
difficulty can be overcome when considering turn-based games, or when
the number of available moves is fixed.  The satisfiability problem
for \ATLsc is then reduced to the satisfiability problem for \QCTL,
which we proved decidable (with non-elementary complexity)
in~\cite{LM13}. When restricting to memoryless strategies, a~similar
reduction to \QCTL exists, but in a setting where the quantified
atomic propositions directly label the model, instead of its
unwinding. The satisfiability problem for \QCTL under that semantics
is undecidable~\cite{Fre01,LM13}, and we adapt the proof of that
result to show that satisfiability of \ATLsc0 (in which quantification
is restricted to memoryless strategies) is also undecidable.

\section{Definitions}
\label{sec-def}

\subsection{\ATL with strategy contexts}


In this section, we~define the framework of concurrent game
structures, and define the logic \ATL with strategy contexts.
We fix once and for all a set~$\AP$ of atomic propositions.

\begin{definition}
A \emph{Kripke structure}~$\calS$ is a $3$-tuple
$\tuple{Q,R,\ell}$ where $Q$~is a countable set of states,
$R \subseteq Q^2$ is a total 
relation 
(\emph{i.e.}, for all~$q\in Q$,
  there is~$q'\in Q$ s.t. $(q,q')\in R$)
and $\ell\colon Q
\rightarrow 2^\AP$ is a labelling function.
\end{definition}

A~path in a Kripke structure~$\calS$ is a mapping~$\rho\colon \mathbb N\to Q$
such that $(\rho(i),\rho(i+1))\in R$ for all~$i$.  We~write $\first\rho=\rho(0)$.  
Given a path~$\rho$ and an
integer~$i$, 
the $i$-th suffix of~$\rho$, is the path~$\rho_{\geq i}\colon n \mapsto \rho(i+n)$;
the $i$-th prefix of~$\rho$, denoted~$\rho_{\leq i}$, is
the finite sequence made of the $i+1$ first state of~$\rho$.  We~write
$\Execf(q)$ for the set of finite prefixes of paths (or
\emph{histories}) with first state~$q$.  We~write $\last\pi$ for the 
last state of a history~$\pi$. Given a
history~$\rho_{\leq i}$ and a path~$\pi$ such that $\last{\rho_{\leq
    i}}=\first\pi$, the concatenation $\lambda=\rho_{\leq i}\cdot\pi$
  is defined by $\lambda(j)=\rho(j)$ when $j\leq i$ and
  $\lambda(j)=\pi(j-i)$ when~$j>i$.

\begin{definition}[\cite{AHK02}]  \label{def-cgs} 
A \emph{Concurrent Game Structure}~(\emph{\CGS{}})~$\mathcal C$ is a
$7$-tuple $\tuple{Q, R, \ell, \Agt, \Alac, \Chc, \Edg}$ where:
$\tuple{Q,R,\ell}$ is a (possibly infinite-state) Kripke structure,
$\Agt=\{a_1,\ldots,a_p\}$ is a finite set of \emph{agents}, \Alac is a
non-empty set of moves, $\Chc\colon Q \times\Agt
\to \Part(\Alac)\smallsetminus\{\varnothing\}$ defines the set of
available moves of each agent in each state, and $\Edg\colon Q \times
\Alac^\Agt \to R$ is a transition table associating, with each
state~$q$ and each set of moves of the agents, the resulting
transition departing from~$q$.
\end{definition}

The size of a \CGS $\calC$ is $|Q|+|\Edg|$.  For a state~$q\in Q$,
we~write $\Next(q)$ for the set of all states reachable by  the 
possible moves from~$q$, and $\Next(q,a_j,m_j)$, with
$m_j\in\Chc(q,a_j)$, for the restriction of $\Next(q)$ to possible
transitions from~$q$ when player~$a_j$ plays move~$m_j$.
We~extend~$\Chc$ and~$\Next$ to coalitions (\ie,~sets of agents) in
the natural~way.
We say that a \CGS is \emph{turn-based} when each state~$q$ is controlled by a
given agent, called the owner of~$q$ (and denoted~$\Own(q)$). In~other terms,
for every~$q\in Q$, for any two move vectors $m$ and~$m'$ in which $\Own(q)$ plays the same move,
it~holds $\Edg(q,m)=\Edg(q,m')$ (which can be
achieved by letting the~sets~$\Chc(q,a)$ be singletons for every $a\not=
\Own(q)$).

A~\newdef{strategy} for some player $a_i\in \Agt$ in a \CGS~$\calC$ is a
function~$f_{i}$ that maps any history to a possible move for~$a_i$,
\ie, satisfying $f_{i}(\pi) \in \Chc(\last\pi,a_i)$.  
A~strategy~$f_i$ is \emph{memoryless} if $f_i(\pi)=f_i(\pi')$ whenever
$\last\pi=\last{\pi'}$. 
A~strategy for a
coalition~$A$ is a mapping assigning a strategy to each agent
in~$A$. The set of strategies for~$A$ is denoted~$\Strat(A)$.
The~\emph{domain}~$\dom(F_A)$ of $F_A\in \Strat(A)$ is~$A$.  Given a
coalition~$B$, the~strategy $(F_{A})_{|B}$
(resp.~$(F_{A})_{\smallsetminus B}$) denotes the restriction of~$F_A$
to the coalition~$A\cap B$ (resp.~$A \smallsetminus B$).  Given two
strategies $F\in\Strat(A)$ and $F'\in\Strat(B)$, we~define $F\compo
F'\in \Strat(A\union B)$ as $(F \compo
F')_{|a_j}(\rho)=F_{|a_j}(\rho)$ (resp.\ $F'_{|a_j}(\rho)$) if $a_j\in
A$ (resp.\ $a_j\in B\smallsetminus A$).

Let~$\rho$ be a history.  A~strategy~$F_A=(f_{j})_{a_j\in A}$ for some
coalition~$A$ induces a set of paths from~$\rho$, called the
\emph{outcomes} of~$F_A$ after~$\rho$, and denoted $\Out(\rho,F_A)$:
an~infinite path $\pi=\rho\cdot q_1 q_2\ldots$ is in~$\Out(\rho,F_A)$
if, and only~if, writing $q_0=\last\rho$, for all~$i\geq 0$ there is a set of
moves~$(m^i_{k})_{a_k\in\Agt}$ such that $m^i_{k}\in\Chc(q_i,a_k)$ for
all~$a_k\in\Agt$, $m^i_{k}=f_{k}(\pi_{|\rho|+i})$ if~$a_k\in A$, and
$q_{i+1}$ is the unique element of   $\Next(q_i,\Agt,(m^i_k)_{a_k\in \Agt})$.
%
Also, given a history~$\rho$ and a strategy~$F_A=(f_{j})_{a_j\in A}$,
the strategy $F_A^{\rho}$ is the sequence of strategies $(f^{\rho}_{j})_{a_j\in A}$  
such that $f_j^{\rho}(\pi) = f_j(\rho\cdot \pi)$, assuming $\last{\rho}=\first{\pi}$.

We now introduce the extension of \ATL with strategy
contexts~\cite{BDLM09,DLM10}:
\begin{definition}\label{def-ATLSes}
Given a set of atomic propositions $\AP$ and a set of agents $\Agt$, the syntax of~\ATLsc* is defined as follows
(where $p$ ranges over~$AP$ and $A$ over~$2^\Agt$):
\begin{xalignat*}1
\ATLsc* \ni \phis,\psis \coloncolonequals & p \mid \non\phis \mid \phis\ou\psis
  \mid   \Relax[A] \phis   \mid   \Diams[A] \phip \\
  \phip,\psip   \coloncolonequals &
  \phis \mid \non\phip \mid \phip\ou\psip \mid \X\phip\mid  \phip\Until\psip. 
\end{xalignat*}
\end{definition}

That a (state or path) formula~$\phi$ is satisfied at a position~$i$ of a path~$\rho$ of a
\CGS~$\calC$ under a strategy context~$F\in\Strat(B)$ (for~some coalition~$B$),
denoted~$\calC, \rho, i \sat_{F}\phi$, is defined as
follows (omitting atomic propositions and Boolean operators):
\begin{xalignat*}1
  \calC, \rho, i \sat_{F}  \Relax[A] \phis &\quad\mbox{ iff }\quad 
  \calC, \rho,  i \sat_{F_{\smallsetminus A}} \phis \\
\calC, \rho, i \sat_{F}  \Diams[A] \phip & \quad\mbox{ iff }\quad   
  \exists F_A\in\Strat(A). \
  \forall \rho'\in\Out(\rho_{\leq i},F_A\compo F).\ 
   \calC, \rho',i \sat_{F_A\compo F} \phip \\
\calC, \rho,i \sat_{F}  \X \phip &\quad\mbox{ iff }\quad    \calC, \rho,i+1{}
 \sat_{F} \phip \\ 
\calC, \rho,i  \sat_{F}  \phip \Until \psip & \quad\mbox{ iff }\quad
 \exists j\geq 0.\
  \calC, \rho, i+j \sat_{F} \psip  \mbox{ and }
\forall 0\leq k < j.\ \calC, \rho, i+k \sat_{F} \phip
\end{xalignat*}
Notice how the (existential) strategy quantifier contains an implicit
universal quantification over the set of outcomes of the selected strategies.
Also notice that state formulas do not really depend on the selected path: indeed 
one can easily show that
%
\[
\calC, \rho, i \sat_F \phis \quad\mbox{ iff }\quad 
 \calC, \rho', j \sat_{F'} \phis  
\]
where we assume $\rho(i)=\rho'(j)$ and where 
$F$ and $F'$ verifies: $F(\rho_{\leq i}\cdot\rho'') = F'(\rho'_{\leq
  j}\cdot\rho'')$ for any finite $\rho''$ starting in~$\rho(i)$. 
In particular this is the case when the $\rho_{\leq i}=\rho'_{\leq j}$ and ${F=F'}$. 

In the sequel we~equivalently write $\calC,\pi(0)\models_F\phis$ 
in place of $\calC,\pi,0\models_F\phis$ when dealing with state formulas.

For convenience, in the following we allow the construct $\Diams[A]
\phis$, defining it as a shorthand for $\Diams[A] \bot \Until
\phis$. We~also use the classical modalities~$\F$ and~$\G$, which can
be defined using~$\Until$. Also, $\Boxs[A] \phip =
\non\Diams[A]\non\phip$ expresses that any $A$-strategy has at least one outcome
where $\phip$~holds. 

The fragment \ATLsc of \ATLsc* is defined as usual, by restricting the
set of path formulas to 
\[
  \phip,\psip   \coloncolonequals  \non\phip \mid \X\phis \mid  \phis\Until\psis. 
\]
It~was proved in~\cite{BDLM09} that \ATLsc is actually as expressive
as \ATLsc*. Moreover, for any given set of players, any \ATLsc formula
can be written without using negation in path formulas, replacing for
instance $\Diams[A]\G\phi$ with
$\Diams[A]\non\Diams[\Agt\setminus(A\cup B)]\F\non\phi$, where $B$ is
the domain of the context in which the formula is being
evaluated. While this is not a generic equivalence (it~depends on the
context and on the set of agents), it~provides a way of removing
negation from any given \ATLsc formula.

\subsection{Quantified \CTL} 

In this section, we introduce \QCTL, and define its \emph{tree
  semantics}.
\begin{definition}
Let~$\Sigma$ be a finite alphabet, and~$S$ be a (possibly infinite)
set of directions.  A~\emph{\mbox{$\Sigma$-labelled} \mbox{$S$-tree}} is a
pair~$\calT = \tuple{T, l}$, where $T\subseteq S^*$ is a non-empty set
of finite words on~$S$ s.t.  for~any non-empty word~$n=m\cdot s$
in~$T$ with~$m\in S^*$ and~$s\in S$, the word~$m$ is also in~$T$; and
$l\colon T \to \Sigma$ is a labelling function.
\end{definition}

The \emph{unwinding} (or \emph{execution tree}) of a Kripke structure
$\calS = \tuple{Q,R,\ell}$ from a state $q\in Q$ is the
$2^\AP$-labelled $Q$-tree $\calT_\calS(q) =
\tuple{\Execf(q),\ell_\calT}$ with $\ell_\calT(q_0 \cdots
q_i)=\ell(q_i)$. Note that
$\calT_\calS(q)=\tuple{\Execf(q),\ell_\calT}$ can be seen as an
(infinite-state) Kripke structure where the set of states is
$\Execf(q)$, labelled according to~$\ell_{\calT}$, and with
transitions $(m,m\cdot s)$ for all~$m\in \Execf(q)$ and $s\in Q$
s.t. $m\cdot s\in \Execf(q)$.

\begin{definition}
For $P\subseteq \AP$, two $2^{\AP}$-labelled trees $\calT=\tuple{T,\ell}$
and $\calT'=\tuple{T',\ell'}$ are \emph{$P$-equivalent} (denoted
by $\calT \equiv_P \calT'$) 
whenever $T=T'$, and $\ell(n)\cap P=\ell'(n)\cap P$ for any~$n\in T$.
\end{definition}

In other terms, $\calT\equiv_P \calT'$ if $\calT'$ can be obtained
from~$\calT$ by modifying the labelling function of~$\calT$ for propositions not in~$P$.
%
%
We now define the syntax and semantics of \QCTL*:
\begin{definition}
\label{def-QCTLs}
The syntax of \QCTL* is defined by the following grammar:
\begin{xalignat*}1
\QCTL*\ni\phis,\psis \coloncolonequals & p \mid \non\phis \mid \phis\ou\psis
       \mid  \Ex \phip  \mid  \All  \phip  \mid  \exists p.\ \phis   \\
 \phip,\psip \coloncolonequals &  \phis \mid \non\phip \mid \phip \ou \psip \mid 
        \X\phip\mid  \phip\Until\psip.
\end{xalignat*}
%
\end{definition}

\QCTL* is interpreted here over Kripke structures through their
unwindings\footnote{Note that several semantics are possible for
  \QCTL* and the one we use here is usually called the \emph{tree
    semantics}.}: given a Kripke structure~$\calS$, a~state~$q$ and a
formula~$\phi\in\QCTL*$, that $\phi$~holds at~$q$ in~$\calS$, denoted
with $\calS,q\models_t \phi$, is defined by the truth value of
$\calT_\calS(q) \sat \phi$ that uses the standard inductive semantics
of \CTL* over trees extended with the following case:
\[
\calT \sat \exists p. \phis  \quad\text{ iff }\quad \exists
\calT' \equiv_{\AP\backslash\{p\}} \calT \text{ s.t. } \calT' \sat \phis. 
\]
Universal quantification over atomic propositions, denoted with the construct
$\forall p.\ \phi$, is obtained by dualising this definition. We~refer
to~\cite{LM13} for a detailed study of~\QCTL* and \QCTL. Here we just recall
the following important properties of these logics. First note that \QCTL is
actually as expressive as \QCTL* (with an effective
translation)~\cite{Fre01,DLM12}. Secondly model checking and satisfiability
are decidable but non elementary. More precisely given a \QCTL formula $\phi$
and a (finite) set of degrees $\calD\subseteq \mathbb{N}$, one can build a
tree automaton $\calA_{\phi,\calD}$ recognizing the $\calD$-trees satisfying
$\phi$. This provides a decision procedure for model checking as the Kripke
structure $\calS$ fixes the set $\calD$, and it remains to check whether the
unwinding of $\calS$ is accepted by $\calA_{\phi,\calD}$. For satisfiability
the decision procedure is obtained by building a formula $\phi_2$ from $\phi$
such that $\phi_2$ is satisfied by some $\{1,2\}$-tree iff $\phi$ is satisfied
by some finitely-branching tree. Finally it remains to notice that a \QCTL
formula is satisfiable iff it is satisfiable in a finitely-branching tree (as
\QCTL is as expressive as~\MSO) to get the decision procedure for \QCTL
satisfiability. By consequence we also have that a \QCTL formula is
satisfiable iff it is satisfied by a regular tree (corresponding to the
unwinding of some finite Kripke structure).

\section{From \ATLsc to \QCTL}

\looseness=-1
The main results of this paper concern the satisfiability problem for
\ATLsc: given a formula in \ATLsc, does there exists a \CGS~$\calC$
and a state~$q$ such that $\calC, q\models_\emptyset \phi$ (with empty initial
context)? Before we present these results in the next sections,
we~briefly explain how we reduce the model-checking problem for \ATLsc
(which consists in deciding whether a given state~$q$ of a given
\CGS~$\calC$ satisfies a given \ATLsc formula~$\phi$) to the
model-checking problem for~\QCTL. This reduction will serve as a basis for
proving our main result.

\subsection{Model checking}
Let $\calC=\tuple{Q, R, \ell, \Agt, \Alac, \Chc, \Edg}$ be a
finite-state \CGS,
with a finite set of moves $\Alac=\{m_1,\ldots,m_k\}$. We~consider the following
sets of fresh
atomic propositions: $\APQ \eqdef \{\psf_q \mid q \in
Q\}$, $\APM{j} \eqdef \{\msf_1^j,\ldots,\msf_k^j\}$ for every
$a_j\in\Agt$, and write $\APMt \eqdef \bigcup_{a_j\in\Agt}\APM{j}$.
Let $\calS_\calC$ be the Kripke structure $\tuple{Q,R,\ell_{+}}$ where
for any state~$q$, we have: $\ell_+(q) \eqdef \ell(q) \cup
\{\psf_q\}$. 
%
A strategy for an agent~$a_j$ can be seen as a function $f_j\colon \Execf(q)
\rightarrow \APM{j}$ labeling the execution
tree of~$\calS_{\calC}$ with propositions in~$\APM{j}$. 

Let $F\in\Strat(C)$ be a strategy context 
and $\Phi\in \ATLsc$. We~reduce the question whether
$\calC, q \sat_F \Phi$ to a model-checking instance for \QCTL*
over~$\calS_{\calC}$. For this, we~define a \QCTL* formula
$\overline{\Phi}^C$ inductively: for non-temporal formulas,
\begin{xalignat*}4
\overline{\Relax[A] \phi}^C  &\eqdef  \overline{\phi\vphantom{\psi}}^{C\smallsetminus A} 
&
\overline{\phi \et \psi}^C &\eqdef  \overline{\vphantom{\psi}\phi}^C  \et
\overline{\vphantom{\psi}\psi}^C  &
\overline{\non \psi}^C  &\eqdef    \non \overline{\vphantom{\psi}\phi}^C 
&
\overline{p}^C  &\eqdef p 
\end{xalignat*}
For a formula of the form $\Diams[A] \X \phi$ with $A=\{a_{j_1},\ldots,a_{j_l}\}$, we let: 
\[
   \overline{\Diams[A] \X \phi}^{\!\!C} \eqdef 
\exists \msf_1^{j_1} ... \msf_k^{j_1} ... \msf_1^{j_l} ...
\msf_k^{j_l}. 
\!\ET_{a_j \in A}\! \AG \Big( \Phistrat(a_j) 
\Big) \et 
\All \Big( \Phiout{C\cup A}  \impl  \X \overline{\phi}^{C\cup A}
\Big)
\]
where:
\begin{xalignat*}1
\Phistrat(a_j) &\eqdef \OU_{q\in Q} \Big( \psf_q \et \OU_{m_i \in
  \Chc(q,a_j)} (\msf_i^j \et \ET_{l\not= i} \non \msf_l^j) \Big) 
\\[1mm]
\Phiout{A} &\eqdef \G \Biggl[
\ET_{\substack{q\in Q \\ m\in \Chc(q,A)}}
\Bigl( (\psf_q \et m) \impl  \X \bigl( \OU_{q' \in \Next(q,A,m)}
  \psf_{q'} \bigr) \Bigr) \Biggr]
\end{xalignat*}
\noindent
where $m$ is a move $(m^j)_{a_j\in A} \in \Chc(q,A)$ for $A$ and $P_m$ is
the propositional formula $\ET_{a_j\in A} m^j$ characterizing~$m$.
Formula~$\Phistrat(a_j)$ ensures that the labelling of propositions~$\msf_i^j$ describes a
feasible strategy for~$a_j$.
Formula~$\Phiout{A}$ characterizes the outcomes
of the strategy for~$A$ that is described by the atomic propositions
in the model. 
Note that $\Phiout{A}$ is based on the transition table~$\Edg$ of~$\calC$ (via~$\Next(q,A,m)$).
For a formula of the form $\Diams[A]  (\phi \Until \psi)$ with $A=\{a_{j_1},\ldots,a_{j_l}\}$, 
we~let: 
\[
   \overline{\Diams[A] (\phi \Until \psi)}^{\!\!C} \eqdef 
\exists \msf_1^{j_1} ... \msf_k^{j_1} ... \msf_1^{j_l} ...
\msf_k^{j_l}. 
\!\ET_{a_j \in A}\! \AG \Big( \Phistrat(a_j) 
\Big) \et 
\All \Big( \Phiout{C\cup A}  \impl  (\overline{\phi}^{C\cup A} \Until \overline{\psi}^{C\cup A})
\Big)
\]
Then:
\begin{restatable}{theorem}{thmatlsctoqctl}\cite{DLM12}
\label{thm-atlsctoqctl}
\label{th:atlsc2qctl}
  Let $q$ be a state in a \CGS~$\calC$.  Let $\Phi$ be an \ATLsc formula and
  $F$ be a strategy context for some coalition~$C$.  Let~$\calT'$ be
  the execution tree $\calT_{\calS_\calC}(q)$ with a labelling function
  $\ell'$ s.t.\ for every $\pi \in \Execf(q)$ of length~$i$ and any $a_j \in C$,
  $\ell'(\pi)\cap \APM{j} = \msf^j_i$ if, and only~if, $F(\pi)_{|a_j}=m_i$. Then
$\calC, q \sat_F \Phi$ if, and only~if,
$\calT', q \sat_t  \overline{\Phi}^C$.
\end{restatable}
Combined with the (non-elementary) decision procedure for \QCTL* model
checking, we~get a model-checking algorithm for model checking \ATLsc. Notice
that our reduction above is into \QCTL*, but as explained before
every \QCTL* formula can be translated into \QCTL. 
%
Finally note that model checking is non elementary (\EXPTIME[k]-hard for
any~$k$) both for \QCTL and \ATLsc~\cite{DLM12}.

%
%

\subsection{Satisfiability}

We now turn to satisfiability.  The reduction to \QCTL 
we just developed for model checking does not extend to satisfiability,
because the \QCTL formula  we built depends both on the formula and on the
structure. 
Actually, satisfiability is undecidable for \ATLsc, both
for infinite \CGS and when restricting to finite \CGS~\cite{TW12}.
It~is worth noticing that both problems are relevant, as \ATLsc does not have
the finite-model property (nor~does it have the finite-branching property).
This can be derived from the fact that the modal logic~$S5^n$ does not have the
finite-model property~\cite{Kur02}, and from the elegant reduction of satisfiability
of~$S5^n$ to satisfiability of~$\ATLsc$ given in~\cite{TW12}~\footnote{Indeed the  finite-branching   property for $\ATLsc$ would imply the finite-model property for $S5^n$.}.


%

In~what follows, we~prove decidability of satisfiability in two different
settings: first in the setting of turn-based games, and  then in the setting
of a bounded number of actions allowed to the players. A~consequence of our
decidability proofs is that in both cases (based on automata constructions), \ATLsc does have the finite-model
property (thanks to Rabin's regularity theorem). We~also consider the setting
where quantification is restricted to memoryless strategies, but prove that
then satisfiability is undecidable (even on turn-based games and with a fixed
number of actions).

Before we proceed to the algorithms for satisfiability, we~prove a
generic result~\footnote{Note that it still holds true when restricting to turn-based games.} about the number of agents needed in a \CGS to satisfy
a formula involving a given set of agents. This result has already been proved for \ATL (\eg in~\cite{jlc16(6)-WLWW}). 
Given a formula $\Phi \in \ATLsc$, we use $\Agt(\Phi)$ to denote the
set of agents involved in the strategy quantifiers in~$\Phi$.


%
\begin{proposition}\label{prop-agt}
An \ATLsc formula~$\Phi$ is satisfiable iff, it is
satisfiable in a \CGS with $|\Agt(\Phi)|+1$ agents.
\end{proposition} 

\begin{proof}
  Assume $\Phi$ is satisfied in a \CGS $\calC=\tuple{Q, R, \ell, \Agt, \Alac,
    \Chc, \Edg}$. If $|\Agt|\leq\Agt(\Phi)$, one can easily add extra players
  in~$\calC$ in such a way that they play no role in the behavior of the game
  structure. Otherwise, if $|\Agt|>\Agt(\Phi)+1$, we can replace the agents
  in~$\Agt$ that do not belong to $\Agt(\Phi)$ by a unique agent mimicking the
  action of the removed players. For example, a~coalition $A =
  \{a_1,\ldots,a_k\}$ can be replaced by a player~$a$ whose moves are
  $k$-tuples in~$\Alac^k$.
\end{proof}





\section{Turn-based case}
\label{sec-tb}

Let $\Phi$ be an \ATLsc formula, and assume $\Agt(\Phi)$ is the set
$\{a_1,\ldots,a_n\}$. Following Prop.~\ref{prop-agt}, let~$\Agt$ be the set of
agents $\Agt(\Phi)\cup \{a_0\}$, where $a_0$ is an additional player. In the
following, we~use an atomic propositions~$(\turnsf_j)_{a_j\in\Agt}$ to specify 
the owner of the states.
A~strategy for an agent~$a_j$ can be encoded
by an atomic proposition $\movsf_j$: indeed it is sufficient to mark one
\emph{successor} of every $a_j$-state (notice that this is a crucial difference
with~\CGS). The outcomes of such a strategy are the runs in which every
$a_j$-state is followed by a state labelled with~$\movsf_j$; this is the main
idea of the reduction below.

Given a coalition~$C$ (which we intend to represent the agents that have a
strategy in the current context), we define a \QCTL* formula
$\widehat{\Phi}^C$ inductively:
\begin{itemize}
\item for non-temporal formulas we let:
\begin{xalignat*}4
\widehat{\Relax[A] \phi}^C  &\eqdef  \widehat{\phi\vphantom{\psi}}^{C\smallsetminus A} 
&
\widehat{\phi \et \psi}^C &\eqdef  \widehat{\vphantom{\psi}\phi}^C  \et
\widehat{\vphantom{\psi}\psi}^C  &
\widehat{\non \psi}^C  &\eqdef    \non \widehat{\vphantom{\psi}\phi}^C 
&
\widehat{P}^C  &\eqdef  P 
\end{xalignat*}
\item for path formulas, we define: 
\begin{xalignat*}2
\widehat{\X \phi}^{C} & \eqdef \X \widehat{\phi}^{C} &
\widehat{\phi \Until \psi}^{C} &\eqdef  \widehat{\phi}^{C} \Until \widehat{\psi}^{C}
\end{xalignat*}
\item for formulas of the form $\Diams[A] \phi$ with $A=\{a_{j_1},\ldots,a_{j_l}\}$, we~let: 
\begin{multline*}
   \widehat{\Diams[A] \phi}^{C} \eqdef 
\exists \movsf_{j_1} ... \movsf_{j_l}. \\
\biggl[ \AG \ET_{a_j\in A} (\turnsf_j \impl \EX[1] \movsf_j) \et   
  \All \Bigl[ \G \Bigl( \ET_{a_j \in A\cup C} (\turnsf_j \impl \X \movsf_j)\Bigr) 
   \thn \widehat{\phi}^{C\cup A} \Bigr] \biggr]
\end{multline*}
where $\EX[1] \alpha$ is a shorthand for
\(
\EX \alpha \et \forall p. \Big( \EX (\alpha \et p) \thn \AX (\alpha \thn p) \Big)
\),
specifying the~existence of a unique successor satisfying~$\alpha$.
\end{itemize}

Now we have the following proposition, whose proof is done by structural 
induction over the formula:
%
\begin{restatable}{proposition}{propsattb}
\label{prop-sat-tb}
Let~$\Phi\in\ATLsc$, and $\Agt=\Agt(\Phi)\cup\{a_0\}$ as above.
Let~$\calC$ be a turn-based \CGS,
$q$~be a state of~$\calC$, and $F$ be a strategy context. Let $\calT_\calC(q)=\tuple{T,\ell}$ be
the execution tree of the underlying Kripke structure of~$\calC$ (including a
labelling with propositions~$(\turnsf_j)_{a_j\in\Agt}$).
Let $\ell_F$ be the labelling extending~$\ell$ such that
for every node $\rho$ of~$T$ belonging to some $a_j \in \dom(F)$ (\ie, such
that $\last\rho\in\Own(a_j)$), its
successor $\rho\cdot q$ according to~$F$ (\ie, such that $F_j(\rho)=q$) is labelled
with~$\movsf_j$.
Then we have:
\[
\calC,q \sat_F \Phi \quad\mbox{iff}\quad 
\tuple{T,\ell_F} \sat \widehat{\Phi}^{\dom(F)} 
\]
\end{restatable}


\begin{proof}
The proof is by structural induction over~$\Phi$. 
The cases of atomic propositions and Boolean operators are straightforward. 
\begin{itemize}
\item $\Phi = \Diams[A] (\phi \Until \psi)$: assume $\calC,q
  \sat_F \Phi$. Then there exists $F_A \in \Strat(A)$ s.t.\ for any
  $\rho\in \Out(q,F_A\compo F)$, there exists $i\geq 0$
  s.t.\ $\calC,\rho(i) \sat_{(F_A\compo F)^{\rho_{\leq i}}} \psi$ and
  $\forall 0\leq j < i$, we have $\calC,\rho(j) \sat_{(F_A\compo
    F)^{\rho_{\leq j}}} \phi$. Let~$\ell_{F_A\compo F}$ be the extension
  of~$\ell$ labelling~$T$ with propositions $(\movsf_j)_{a_j\in \Agt}$
  according to the strategy context~$F_A\compo F$.
  By induction hypothesis, the following two
  statements hold true:
\begin{itemize}
\item $\tuple{T,\ell_{F_A\compo F}}_{\rho_{\leq i}} \sat
    \widehat{\psi}^{\dom(F)\cup A}$, and
  \item $\tuple{T,\ell_{F_A\compo F}}_{\rho_{\leq j}} \sat
      \widehat{\phi}^{\dom(F)\cup A}$ for any $0\leq j < i$.
\end{itemize}
(where $\tuple{U,l}_{\pi}$ is the subtree of~$\tuple{U,l}$ rooted at
  node~$\pi\in U$).
As this is true for every~$\rho$ in the outcomes induced by $F_A\compo
F$, it~holds for every path in the execution tree satisfying the
constraint over the labelling of $(\turnsf_j)_{a_j\in\Agt}$
and~$(\movsf_j)_{a_j\in\Agt}$. It~follows that
\[
\tuple{T,\ell_{F_A\compo F}} \sat  \All \Big[ \G \Big( 
  \ET_{a_j \in A\cup C} (\turnsf_j \impl \X \movsf_j)\Big)
 \thn \widehat{\phi}^{\dom(F)\cup A} \Big] 
\]

Moreover we also know that $\AG \ET_{a_j\in A} (\turnsf_j \impl \EX_1 \movsf_j)$
holds true in~$\tuple{T,\ell_{F_A\compo F}}$ since the
labelling~$\ell_{F_A\compo F}$ includes the strategy~$F_A$.
Hence $\tuple{T,\ell_F} \sat \widehat{\Phi}^{\dom(F)}$, with the labelling 
for $(\movsf_j)_{a_j\in A}$ being obtained from~$F_A$.

\smallskip
Now assume $\tuple{T,\ell_F} \sat \widehat{\Phi}^{\dom(F)}$. 
Write $A=\{a_{j_1},\ldots,a{j_l}\}$. Then we have:
\begin{multline*}
  \tuple{T,\ell_F} \sat \exists \movsf_{j_1} ... \movsf_{j_l}.  
    \Bigl[ \AG \ET_{a_j\in A} (\turnsf_j \impl \EX_1 \movsf_j) \et  \\
  \All \Bigl[ \G \Bigl( \ET_{a_j \in A\cup C} (\turnsf_j \impl \X \movsf_j)\Bigr)
  \thn (\widehat{\phi}^{\dom(F)\cup A} \Until \widehat{\psi}^{\dom(F)\cup
    A})\Bigr] \Bigr]
\end{multline*}
The first part of the formula, namely  $\AG \ET_{a_j\in A} (\turnsf_j \impl \EX_1
\movsf_j)$, ensures that the labeling with~$(\movsf_j)_{a_j\in A}$ defines a 
strategy for the coalition~$A$. The second part
states that every run belonging to the outcomes of $F_A\compo F$ (remember
that $\ell_F$ already contains the strategy context~$F$) satisfies
$(\widehat{\phi}^{\dom(F)\cup A} \Until \widehat{\psi}^{\dom(F)\cup A})$.
Finally it remains to use the induction hypothesis over states along the
execution to deduce $\calC,q \sat_F \Diams[A] (\phi \Until \psi)$.

\item $\Phi = \Relax[A] \psi$:  assume $\calC,q
  \sat_F \Phi$. Then $\calC,q\models_{F_{\dom(F)\setminus A}}
  \psi$. Applying the
  induction hypothesis, we~get $\tuple{T,\ell_{F_{\dom(F)\setminus A}}} \sat
  \widehat{\psi}^{{\dom(F)\setminus A}}$. And
  it~follows that $\tuple{T,\ell_{F}} \sat \widehat{\psi}^{{\dom(F)\setminus A}}$ because the labeling of strategies for coalition $A$ in $F$ is not used for evaluating $\widehat{\psi}^{{\dom(F)\setminus A}}$. 
Conversely, assume $\tuple{T,\ell_{F}} \sat \widehat{\psi}^{{\dom(F)\setminus A}}$. Then we have 
$\tuple{T,\ell_{F_{\dom(F)\setminus A}}} \sat \widehat{\psi}^{{\dom(F)\setminus A}}$ (again the labeling of $A$ strategies in $F$ is not used for evaluating the formula). Applying induction hypothesis, we get $\calC,q\models_{F_{\dom(F)\setminus A}} \psi$ and then  $\calC,q
  \sat_F \Phi$.


\item $\Phi \eqdef \Diams[A] \X \phi$ and $\Phi= \Relax[A] \X\phi$: the proofs are similar to the
  previous~ones.\qed
\end{itemize}
\let\qed\relax
\end{proof}


Finally, let $\Phi_{tb}$ be the following formula, 
used to make the game turn-based:
\[
\Phi_{tb} = \AG \Big[ \OU_{a_j \in \Agt} \Big( \turnsf_j \et \ET_{a_l\not= a_j} \non \turnsf_{l} \Big) \Big]
\]
and let $\widetilde{\Phi}$ be the formula $\Phi_{tb} \et \widehat{\Phi}^\emptyset$. 
Then we have:
\begin{theorem}
\label{th-sat-tb-algo}
Let $\Phi$ be an \ATLsc formula and $\widetilde{\Phi}$ be the \QCTL* formula
defined as above. $\Phi$~is satisfiable in a 
turn-based \CGS if, and only if, $\widetilde{\Phi}$ is satisfiable (in~the tree
semantics).
\end{theorem}

\begin{proof}
If $\Phi$ is satisfiable in a turn-based structure, then there exists
such a structure $\calC$ with $|\Agt(\Phi)|+1$ agents.  Assume
$\calC,q\sat \Phi$.  Now consider the execution tree $\calT_\calC(q)$ with
the additional labelling  to mark states with the correct
propositions~$(\turnsf_j)_{a_j\in\Agt}$, indicating the owner of each state. From
Proposition~\ref{prop-sat-tb}, we have $\calT_\calC(q)\sat
\widehat{\Phi}^\emptyset$. Thus clearly $\calT_\calC(q) \sat
\widetilde{\Phi}$.

Conversely assume $\calT \sat \widetilde{\Phi}$. As explained in Section~\ref{sec-def}, we can assume that $\calT$ is regular. Thus $\calT \sat \Phi_{tb} \et
\widehat{\Phi}^\emptyset$: the first part of the formula ensures that every
state of the underlying Kripke structure can be assigned to a unique agent,
hence defining a turn-based \CGS. The second part ensures that $\Phi$~holds
for the corresponding game, thanks to Proposition~\ref{prop-sat-tb}.
%
\end{proof}

The above translation from \ATLsc into \QCTL* transforms a formula
with~$k$ strategy quantifiers into a formula with at most $k+1$ nested blocks
of quantifiers; satisfiability of a \QCTL* formula with $k+1$ blocks of
quantifiers is in \EXPTIME[(k+3)]~\cite{LM13}. Hence the algorithm has
non-elementary complexity.
%
We now prove that this high complexity cannot be avoided:
\begin{theorem}
\label{th-sat-tb}
Satisfiability of \ATLsc formulas in turn-based \CGS is non-elementary
(\ie, it~is \EXPTIME[k]-hard, for all~$k$).
\end{theorem}

\begin{proof}[Proof (sketch)]
  Model checking \ATLsc over turn-based games is non-elementary~\cite{DLM12},
  and it can easily be encoded as a satisfiability problem. Let
  $\calC=\tuple{Q, R, \ell, \Agt, \Alac, \Chc, \Edg}$ be a turn-based \CGS,
  and $\Phi$ be an \ATLsc formula. Let $\prop_q$ be a fresh atomic proposition
  for every~$q\in Q$. Now we define an \ATLsc formula~$\Psi_{\calC}$ to
  describe the game~$\calC$ as follows:
\begin{multline*}
\Psi_{\calC} = \AG \Bigl(
  \OU_{q\in Q} (\prop_q \et \ET_{q'\not = q} \non\prop_{q'} \et 
           \ET_{P\in\ell(q)} P \et \ET_{P'\not\in \ell(q)} \non P')\Bigr) \et \\
  \AG \Bigl[ \OU_{q\in Q} \Bigl(\prop_q \impl ( \ET_{q\fleche q'} 
   \Diam[\Own(q)] \X \prop_{q'} \et 
  \ET_{q'.\ q\not\fleche q'} \non \Diam[\Own(q)] \X \prop_{q'}) \Bigr) \Bigr]
\end{multline*}
where $q\fleche q'$ denotes the existence of a transition from~$q$ to~$q'$
in~$\calC$. Any~turn-based \CGS satisfying $\Psi_{\calC}$ corresponds to
some unfolding of~$\calC$, and then has the same execution tree. 
Finally we clearly have that $\calC,q\sat \Phi$ if, and only~if,
$\Psi_{\calC} \et \prop_q \et \Phi$ is satisfiable in a turn-based structure.
\end{proof}

\section{Bounded action alphabet}
\label{sec-ba}

We~consider here another setting where the reduction to \QCTL* can be used to
solve the satisfiability of \ATLsc: we~assume that each player has a bounded
number of available actions.
Formally, it~corresponds to the following satisfiability problem:
\problem{$(\Agt,\Alac)$-satisfiability}
  {a finite set of moves~$\Alac$, a~set of agents~$\Agt$, and an \ATLsc formula~$\Phi$ involving the agents in~\Agt}
  {does there exist a \CGS $\calC=\tuple{Q, R, \ell, \Agt, \Alac, \Chc, \Edg}$ and 
    a state $q\in Q$ such that $\calC,q \sat \Phi$.} 

%
  Assume $\Alac = \{1,\ldots,\alpha\}$ and $\Agt = \{a_1,\ldots, a_n\}$.
  With this restriction, we know that we are looking for a \CGS whose
  execution tree has nodes with degrees in the set
  $\calD=\{1,2,\ldots,\alpha^{n}\}$. We~consider such $\calD$-trees
  where the transition table is encoded as follows: for~every agent $a_i$ and
  move~$m$ in~$\Alac$, we~use the atomic proposition $\movsf^m_i$ to specify
  that agent~$a_i$ has played move~$m$ in the \emph{previous} node. Any
  execution tree of such a \CGS satisfies formula
\[
\Phi_{\Edg} =  \AG  
\Bigl[  
\Bigl( \ET_{\bar{m} \in \Alac^{n}} \EX_1
  \movsf^{\bar{m}}\Bigr)
 \et \AX \Bigl( \OU_{\bar{m}\in\Alac^{n}} \movsf^{\bar{m}}  \Bigr) \Bigr]
\]
where $\movsf^{\bar{m}}$ stands for ${\ET_{a_j \in \Agt} \movsf_j^{\bar{m}_j}}$. 
Notice that the second part of the formula is needed because of the way we
handle the \emph{implicit} universal quantification associated with the
strategy quantifiers of~\ATLsc.

Given a coalition~$C$, we~define a \QCTL* formula
$\widetriangle{\Phi}^C$ inductively as follows:
\begin{itemize}
\item for non-temporal formulas we let
\begin{xalignat*}4
\widetriangle{\Relax[A] \phi}^C  &\eqdef  \widetriangle{\phi\vphantom{\psi}}^{C\smallsetminus A} 
&
\widetriangle{\phi \et \psi}^C &\eqdef  \widetriangle{\vphantom{\psi}\phi}^C  \et
\widetriangle{\vphantom{\psi}\psi}^C  &
\widetriangle{\non \psi}^C  &\eqdef    \non \widetriangle{\vphantom{\psi}\phi}^C 
&
\widetriangle{P}^C  &\eqdef  P 
\end{xalignat*}
\item for temporal modalities, we define 
\begin{xalignat*}2
\widetriangle{\X \phi}^{C} &\eqdef \X \widetriangle{\phi}^{C} &
  \widetriangle{\phi \Until \psi}^{C} &\eqdef  \widetriangle{\phi}^{C} \Until
  \widetriangle{\psi}^{C}.
\end{xalignat*}

\item finally, for formulas of the form $\Diams[A] \phi$ with
  $A=\{a_{j_1},\ldots,a_{j_l}\}$,  we~let: 
\begin{multline*}
  \widetriangle{\Diams[A] \phi}^{C} \eqdef \exists \choosesf_{j_1}^{1} \ldots
  \choosesf_{j_1}^\alpha\ldots
  \choosesf_{j_l}^{1} \ldots \choosesf_{j_l}^\alpha.  \\
  \Bigl[ \AG \Bigl(\ET_{a_j\in A}\ \OU_{m=1\ldots \alpha} (\choosesf^m_j \et 
    \ET_{n\not= m} \non\choosesf^n_j) \Bigr)  \et \\
  \All \Bigl[ \G \Bigl( \ET_{a_j \in A\cup C}\ \ET_{m=1\ldots \alpha}
  (\choosesf^m_j \thn \X \movsf_j^m )\Bigr) \thn \widetriangle{\phi}^{C\cup A} \Bigr]
  \Bigr].
\end{multline*}
The first part of this formula requires that the atomic propositions
$\choosesf^{m}_j$ describe a strategy, while the second part expresses that
every execution following the labelled strategies (including those for~$C$)
satisfies the path formula~$\widetriangle{\phi}^{C\cup A}$. 
\end{itemize}

Now, letting $\wideparen{\Phi}$ be the formula $\Phi_{\Edg}\et
\widetriangle{\Phi}^\emptyset$, we~have the following theorem (similar to Theorem~\ref{th-sat-tb-algo}):
%
\begin{restatable}{theorem}{thmsatba}
\label{thm-sat-ba}
  Let $\Phi$ be an \ATLsc formula, $\Agt=\{a_1,\ldots,a_n\}$ be a finite set
  of agents, $\Alac=\{1,\ldots,\alpha\}$ be a finite set of moves, and
  $\wideparen{\Phi}$ be the formula defined above. Then $\Phi$~is
  $(\Agt,\Alac)$-satisfiable in a \CGS if, and only if, the \QCTL* formula
  $\wideparen{\Phi}$ is satisfiable (in~the tree semantics).
\end{restatable}

We~end up with a non-elementary algorithm (in~\EXPTIME[(k+2)] for a
formula involving $k$ strategy quantifiers) for solving satisfiability
of an \ATLsc formula for a bounded number of moves, both for a fixed
or for an unspecified set of agents (we~can infer the set of agents
using Prop.~\ref{prop-agt}). Since \ATLsc model checking is
non-elementary even for a fixed number of moves (the crucial point is
the alternation of strategy quantifiers), we~deduce:
\begin{corollary}
$(\Agt,\Alac)$-satisfiability for \ATLsc formulas is non-elementary
  (\ie, \EXPTIME[k]-hard, for all~$k$).
\end{corollary}

\section{Memoryless strategies}

\looseness=-1
Memoryless strategies are strategies that only depend on the present
state (as~opposed to general strategies, whose values can depend on
the whole history). Restricting strategy quantifiers to memoryless
strategies in the logic makes model checking much easier: in a finite
game, there are only finitely many memoryless strategies to test, and
applying a memoryless strategy just amounts to removing some
transitions in the graph.
Still, quantification over memoryless strategies is not possible in plain \ATLsc, 
and this additional expressive power 
turns out to make satifiability undecidable, even when restricting to turn-based games.
One should notice that the undecidability proof of~\cite{TW12}
for satisfiability in concurrent games uses one-step games (\ie, they
only involve one \X modality), and hence also holds for memoryless
strategies.

\begin{theorem}\label{thm-memoryless}
Satisfiability of \ATLsc0 (with memoryless-strategy quantification) is
undecidable, even when restricting to turn-based games.
\end{theorem}

\begin{proof}
\looseness=-1
We prove the result for infinite-state turn-based games, by adapting
the corresponding proof for~\QCTL under the structure
semantics~\cite{Fre01}, which consists in encoding the problem of tiling
a quadrant. The result for finite-state turn-based games
can be obtained using similar (but more involved) ideas, by encoding
the problem of tiling all finite grids (see~\cite{LM13} for the
corresponding proof for \QCTL). 

We~consider a finite set~$T$ of tiles, and two binary relations $H$
and~$V$ indicating which tile(s) may appear on the right and above
(respectively) a given tile.  Our proof consists in writing a formula
that is satisfiable only on a grid-shaped (turn-based) game
structure representing a tiling of the quadrant (\ie, of~$\mathbb
N\times \mathbb N$). 
The reduction involves two players: Player~$1$ controls square states
(which are labelled with~$\carre$), while Player~$2$ controls circle
states (labelled with~$\rond$).
Each state of the grid is intended to represent one cell of the
quadrant to be tiled. For technical reasons, the reduction is not that
simple, and our game structure will have three kinds of states (see~Fig.~\ref{fig-tiling}): 
\begin{itemize}
\item the ``main'' states (controlled by Player~$2$), which form the grid.
  Each state in this main part has a \emph{right} neighbour and a \emph{top}
  neighbour, which we~assume we~can identify: more precisely, we~make use of 
  two atomic propositions~$v_1$ and~$v_2$ which alternate along the horizontal
  lines of the grid. The \emph{right} successor of a $v_1$-state is labelled with~$v_2$,
  while its \emph{top} successor is labelled with~$v_1$;
\item the ``tile'' states, labelled with one item of~$T$ (seen as
  atomic propositions). Each tile state only has outgoing
  transition(s) to a tile state labelled with the same tile;
\item the ``choice'' states, which appear between ``main'' states and
  ``tile'' states: there is one choice state associated with each main
  state, and each choice state has a transition to each tile
  state. Choice states are controlled by Player~$1$.
\end{itemize}
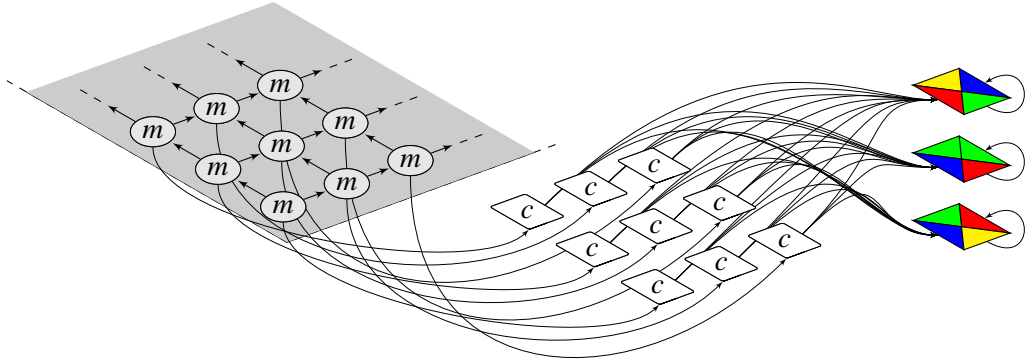
\begin{figure}[!ht]
\centering
\begin{tikzpicture}
\draw[dashed] (20:38mm) -- (0,0) -- (150:43mm);
\draw (20:35mm) -- (0,0) -- (150:40mm);
\fill[black!20!white] (20:35mm) -- (0,0) -- (150:40mm) -- +(20:35mm) -- cycle;
\foreach \x in {5,14,23} 
  {\foreach \y in {6,16,26} 
    {\path (20:\x mm) -- +(150:\y mm) node[coordinate] (d\x\y) {};
     \begin{scope}[xshift=5cm,yshift=-1.3cm]
     \path (20:\x mm) -- +(150:\y mm) node[coordinate] (e\x\y) {};
     \fill[draw,rounded corners=.3mm,fill=black!10!white,opacity=1] 
      (e\x\y) -- ++(20:5mm) node[midway,coordinate] (a\x\y) {} 
          -- ++(150:6mm) node[midway,coordinate] (b\x\y) {} 
          -- ++ (20:-5mm) node[midway,coordinate] (c\x\y) {} 
          -- (e\x\y) node[midway,coordinate] (f\x\y) {} -- (a\x\y);
     \end{scope}
     \draw (d\x\y) node[draw,ellipse,minimum height=4mm,minimum width=6mm,inner
       sep=0pt,fill=white] (d\x\y) {$m$} ;
\path[use as bounding box] (0,0);
     \draw[line width=.1pt,-latex'] (d\x\y) .. controls +(-90:8mm+\x mm) and +(-135:\x mm+5mm) .. (e\x\y);
     \draw[-latex'] (d\x\y) -- +(20:6mm);
     \draw[-latex'] (d\x\y) -- +(150:7mm);
}}
\begin{scope}[xshift=9cm,yshift=-1cm]
\foreach \t/\c/\d/\e/\f in {1/red/green/blue/yellow,2/green/green/blue/red,3/blue/yellow/red/green}
  {\begin{scope}[yshift=\t*9mm,scale=.13]
     \fill[\c,draw=black] (-.25,2.36) -- (4.7,1.71) -- (-.5,4.72) 
       node[midway,coordinate] (DT\t) {} -- cycle;
     \fill[\d,draw=black] (-.25,2.36) -- (-.5,4.72) -- (-5.2,3) -- cycle;
     \fill[\e,draw=black] (-.25,2.36) -- (-5.2,3) -- (0,0)
       node[midway,coordinate] (Tile\t) {} -- cycle;
     \fill[\f,draw=black] (-.25,2.36) -- (0,0) -- (4.7,1.71)
       node[midway,coordinate] (CT\t) {} -- cycle;
     \draw[line width=.1pt,-latex'] (CT\t) {} .. controls +(-30:60mm) and
       +(30:60mm) .. (DT\t);
   \end{scope}}
\foreach \x in {5,14,23} 
  {\foreach \y in {6,16,26} 
    {\foreach \t in {1,2,3}
      {\draw[line width=.1pt,-latex'] (b\x\y) .. controls +(45:40mm-\x mm) and +(-180:10mm) .. (Tile\t);
}}}
\end{scope}
\foreach \x in {5,14,23} 
  {\foreach \y in {6,16,26} 
    {\begin{scope}[xshift=5cm,yshift=-1.3cm]
     \fill[draw,rounded corners=.3mm,fill=white,opacity=1] 
      (e\x\y) -- ++(20:5mm) node[midway,coordinate] (a\x\y) {} 
          -- ++(150:6mm) node[midway,coordinate] (b\x\y) {} 
          -- ++ (20:-5mm) node[midway,coordinate] (c\x\y) {} 
          -- (e\x\y) node[midway,coordinate] (f\x\y) {} -- (a\x\y);
     \end{scope}
     \draw (d\x\y) node[draw,ellipse,minimum height=4mm,minimum width=6mm,inner
       sep=0pt,fill=black!10!white] (d\x\y) {$m$};
     \path (e\x\y) -- +(-.025,.23) node {$c$};
}}
\foreach \x in {5,14,23} 
  {\path (150:34mm) -- ++(20:\x mm) node[coordinate] (a) {};
   \draw [dashed] (a) -- +(150:4mm);}
\foreach \x in {6,16,26} 
  {\path (20:30mm) -- ++(150:\x mm) node[coordinate] (a) {};
   \draw [dashed] (a) -- +(20:4mm);}
\end{tikzpicture}
\caption{The turn-based game encoding the tiling problem}
\label{fig-tiling}
\end{figure}
Assuming that we have such a structure, 
a~tiling of the grid corresponds to a \emph{memoryless} strategy of Player~$1$
(who only plays in the ``choice'' states). Once such a memoryless strategy for Player~$1$ has been selected, that it corresponds to a valid tiling can be
expressed easily: for instance, in any cell of the grid (assumed to be
labelled with~$v_1$), there must exist a
pair of tiles $(t_1,t_2)\in H$ such that 
$v_1 \et \Diamsz[2]\X\X t_1 \et \Diamsz[2]\X(v_2 \et \X\X t_2)$. 
This would be written as follows:
\[
\Diamsz[1] \G\left[
\begin{array}{ll}
&
\displaystyle
v_1 \thn  \OU_{(t_1,t_2)\in H} \Diamsz[2]\X\X t_1 \et \Diamsz[2]\X(v_2 \et \X\X t_2)\\
\et&\\
&
\displaystyle
v_2 \thn  \OU_{(t_1,t_2)\in H} \Diamsz[2]\X\X t_1 \et \Diamsz[2]\X(v_1 \et \X\X t_2)
\end{array}
\right].
\]
The same can be
imposed for vertical constraints, and for imposing a fairness
constraint on the base line (under the same memoryless strategy for Player~$1$). 

\begin{figure}[!ht]
\centering
\begin{minipage}[t]{.45\linewidth}
\centering
\begin{tikzpicture}[scale=1.5]
\begin{scope}
\draw[rounded corners=3mm,dashed,fill=black!10!white] (-.3,0) |- (1.3,1.3) |- (-.3,-.3) -- (-.3,0);
\draw (0,0,0) node[carre] (A1) {};
\draw (1,0,0) node[rond] (B1) {} node {$\scriptstyle\alpha$};
\draw (0,1,0) node[rond] (C1) {} node {$\scriptstyle\beta$};
\draw (1,1,0) node[rond] (D1) {};
\draw (2.3,1.4) node[inner sep=1mm] (C) {to $c$-state};
\draw[-latex'] (D1) .. controls +(60:8mm) and +(-150:6mm) .. (C.-170);
\draw[-latex'] (A1) -- (B1);
\draw[-latex'] (A1) -- (C1);
\draw[-latex'] (B1) -- (D1);
\draw[-latex'] (C1) -- (D1);
\draw (D1) .. controls +(160:5mm) and +(-90:2mm) .. (.5,1.3);
\draw (D1) .. controls +(-70:5mm) and +(-180:2mm) .. (1.3,.5);
\draw[latex'-] (A1) .. controls +(-20:5mm) and +(90:2mm) .. (.5,-.3);
\draw[latex'-] (A1) .. controls +(110:5mm) and +(0:2mm) .. (-.3,.5);
\end{scope}
\draw[dashed] (.5,1.3) -- (.5,1.7);
\draw[dashed] (1.3,.5) -- (1.7,.5);
\draw[dashed] (-.7,.5) -- (-.3,.5);
\draw[dashed] (.5,-.7) -- (.5,-.3);
\end{tikzpicture}
\caption{The cell gadget}
\label{fig-cell}
\end{minipage}
\quad
\begin{minipage}[t]{.45\linewidth}
\centering
\begin{tikzpicture}[scale=.8]
\foreach \x/\y/\h in {0/0/v_1,0/2/v_1,2/0/v_2,2/2/v_2,4/0/v_1,4/2/v_1}
  {\begin{scope}[xshift=\x cm,yshift=\y cm]
\draw[rounded corners=3mm,dashed,fill=black!10!white] (-.3,0) |- (1.3,1.3) |- (-.3,-.3) -- (-.3,0);
\draw (0.1,0.1) node[moycarre] (A1) {} node{$\vphantom{f}\h$};
\draw (.9,.9) node[moyrond] (D1) {};
\begin{scope}[opacity=.5]
\draw (.9,0.1) node[ptrond] (B1) {} node {$\scriptscriptstyle\alpha$};
\draw (0.1,.9) node[ptrond] (C1) {} node {$\scriptscriptstyle\beta$};
\draw[-latex'] (A1) -- (B1);
\draw[-latex'] (A1) -- (C1);
\draw[-latex'] (B1) -- (D1);
\draw[-latex'] (C1) -- (D1);
\end{scope}
\draw (D1) .. controls +(160:5mm) and +(-90:2mm) .. (.5,1.3);
\draw (D1) .. controls +(-70:5mm) and +(-180:2mm) .. (1.3,.5);
\draw[latex'-] (A1) .. controls +(-40:5mm) and +(90:2mm) .. (.5,-.3);
\draw[latex'-] (A1) .. controls +(125:5mm) and +(0:2mm) .. (-.3,.5);
\end{scope}}

\draw (.5,1.3) -- (.5,1.7);
\draw (2.5,1.3) -- (2.5,1.7);
\draw (4.5,1.3) -- (4.5,1.7);
\draw (1.3,.5) -- (1.7,.5);
\draw (1.3,2.5) -- (1.7,2.5);
\draw (3.7,.5) -- (3.3,.5);
\draw (3.7,2.5) -- (3.3,2.5);
\draw[dashed] (-.7,.5) -- (-.3,.5);
\draw[dashed] (-.7,2.5) -- (-.3,2.5);
\draw[dashed] (5.7,.5) -- (5.3,.5);
\draw[dashed] (5.7,2.5) -- (5.3,2.5);
\draw[dashed] (.5,-.7) -- (.5,-.3);
\draw[dashed] (2.5,-.7) -- (2.5,-.3);
\draw[dashed] (4.5,-.7) -- (4.5,-.3);
\draw[dashed] (.5,3.7) -- (.5,3.3);
\draw[dashed] (2.5,3.7) -- (2.5,3.3);
\draw[dashed] (4.5,3.7) -- (4.5,3.3);
\end{tikzpicture}
\caption{Several cells forming (part~of) a grid}
\label{fig-grid}
\end{minipage}
\end{figure}
It~remains to build a formula characterising an infinite grid. This requires a
slight departure from the above description of the grid: each main state will
in fact be a gadget composed of four states, as depicted on
Fig.~\ref{fig-cell}. The first state of each gadget will give the opportunity
to Player~$1$ to \emph{color} the state with either~$\alpha$ or~$\beta$. This
will be used to enforce ``confluence'' of several transitions to the same
state (which we need to express that the two successors of any cell of the
grid share a common successor).

\smallskip 

We now start writing our formula, which we present as a conjunction of several
subformulas. We~require that the main states be labelled with~$m$, the choice
states be labelled with~$c$, and the tile states be labelled with the names of
the tiles. We~let $\AP'=\{m,c\}\cup T$ and $\AP=\AP'\cup\{v_1,v_2,\alpha,\beta,
\carre,\rond\}$. The first part of the formula reads a follows (where
universal path quantification can be encoded, as~long as the context is empty,
using $\Diamsz[\emptyset]$):
\begin{multline}
\All\G\left[\OU_{p\in\AP'} p \et \ET_{p'\in\AP'\setminus\{p\}} \non p'
  \right] 
\et 
\All (m \WUntil c) 
\et 
\All\G\left[c \thn \left(\carre \et \ET_{t\in T} \Diamsz[1]\X t
  \et  \All\X \left(\OU_{t\in T} \All\G t\right)
  \right)\right]
\et\\
\All\G
\left[
(\carre\iff\non\rond) \et 
\left(\carre \thn \ET_{p\in\AP}( \Ex\X p \iff \Diamsz[1]\X p)\right) \et
\left(\rond \thn \ET_{p\in\AP}( \Ex\X p \iff \Diamsz[2]\X p)\right) 
\right]
\label{eq1}
\end{multline}
This formula enforces that each state is labelled with exactly one
proposition from~$\AP'$. It~also enforces that any path will wander
through the main part until it possibly goes to a choice state (this
is expressed as $\All(m\WUntil c)$, where $m\WUntil c$ means $\G m \ou
m\Until c$, and can be expressed a negated-until formula).  Finally,
the second part of the formula enforces the witnessing structures to
be turn-based.

\smallskip
Now we~have to impose that the $m$-part has the shape of a grid:
intuitively, each cell has three successors: one ``to~the right'' and
one ``to~the top'' in the main part of the grid, and one $c$-state
which we will use for associating a tile with this cell.  For
technical reasons, the situation is not that simple, and each cell is
actually represented by the gadget depicted on
Fig.~\ref{fig-cell}. Each state of the gadget is labelled
with~$m$. We~constrain the form of the cells as follows:
\begin{multline}
\All\G\Bigl[m \thn ((\square \et\non\alpha\et\non\beta) 
  \ou (\rond \et \non(\alpha\et\beta)))\Bigr] 
\et 
\All\G\Bigl[
 \bigl( (m\et\carre) \thn (v_1\iff \non v_2)\bigr) \et \bigl((v_1\ou v_2) \thn (m\et\carre)\bigr)
\Bigr] \et
\\
\All\G\Bigl[(m\et\carre) \thn \bigl[\All\X\bigl(m \et \rond \et (\alpha\ou\beta)
  \et \All\X (m \et \rond\et\non\alpha\et\non\beta)\bigr) \et   \Diamsz[1]\X
  \alpha \et \Diamsz[1]\X\beta\bigr]\Bigr]
\end{multline}
This says that there are four types of states in each cell, and
specifies the possible transitions within such cells. 
We now express constraints on the transitions leaving a cell:
\begin{multline}
\All\G\Bigl[(\Ex\X c\ou\Ex\X v_1 \ou \Ex\X v_2) \thn (m \et \rond\et\non\alpha\et\non\beta)\Bigr]
\et \\
\All\G\Bigl[(m \et \rond\et\non\alpha\et\non\beta) \thn (\Ex\X c \et \Ex\X v_1
\et \Ex\X v_2 \et \All\X(c\ou v_1\ou v_2)\Bigr]
\label{eq-leavecell}
\end{multline}

It~remains to enforce that the successor of the $\alpha$ and $\beta$ states
are the same. This is obtained by the following formula:
\begin{equation}
\All\G\bigl[
 (m \et \carre) \thn \Boxsz[2]\bigr(\Diamsz[\emptyset]\X^3 (c \ou v_1) \ou
 \Diamsz[\emptyset]\X^3(c \ou v_2)\bigr)
\bigr]
\label{eq5}
\end{equation}
Indeed, assume that some cell has two different ``final'' states; then there would
exist a strategy for Player~$2$ (consisting in playing differently in those
two final states) that would violate Formula~\eqref{eq5}. Hence each cell as a
single final state.

We~now impose that each cell in the main part
has exactly two $m$-successors, and these two $m$-successors have an
$m$-successor in common. For the former property, Formula~\eqref{eq-leavecell} 
already imposes that each cell has at least two $m$-successors (one labelled 
with~$v_1$ and one with~$v_2$). We~enforce that there cannot be more that two:
\begin{equation}
\All\G\Bigl[(m\et\carre) \thn 
  \Boxsz[1][(\Diamsz[2]\X^3 (v_1\et\X\alpha) \et \Diamsz[2]\X^3(v_2 \et \X\alpha))\thn
  \Boxsz[2]\Diamsz[\emptyset]\X^3\X\alpha] \Bigr].
\label{eq-twosucc}
\end{equation}
Notice that $\Boxsz[2]\Diamsz[\emptyset] \phi$ means that $\phi$~has
to hold along any outcome of any \emph{memoryless} strategy of
Player~$2$.
Assume that a cell has three (or more) successor cells. Then at least
one is labelled with~$v_1$ and at least one is labelled
with~$v_2$. There is a strategy for Player~$1$ to color one
$v_1$-successor cell and one $v_2$-successor cell with~$\alpha$, and a
third successor cell with~$\beta$, thus violating Formula~\eqref{eq-twosucc}
(as Player~$2$ has a strategy to reach a successor cell colored with~$\beta$)

For the latter property (the two successors
have a common successor), we~add the following formula (as well as its $v_2$-counterpart):
\begin{equation}
\Boxsz[1]\Diamsz[\emptyset]\G\Bigl[
  (m\et\carre\et v_1) \thn \Bigl([\Diamsz[2]\X^3(v_1 \et
  \Boxsz[2]\X^3\X\alpha)]
  \thn
  [\Diamsz[2]\X^3(\non v_1 \et \X^3(\non v_1 \et \X\alpha))]\Bigr)
  \Bigr]
\end{equation}
In this formula, the initial (universal) quantification over
strategies of Player~$1$ fixes a color for each cell. The formula
claims that whatever this choice, if we are in some $v_1$-cell and can
move to another $v_1$-cell whose two successors have color~$\alpha$,
then also we can move to a $v_2$-cell having one~$\alpha$
successor (which we require to be a $v_2$-cell). 
As this must hold for any coloring, both successors of the
original $v_1$-cell share a common successor. Notice that this does
not prevent the grid to be collapsed: this would just indicate that
there is a \emph{regular} infinite tiling.

We conclude by
requiring that the initial state be in a square state of a cell in the
main part. 
\end{proof}

\section{Results for Strategy Logic}
\label{sec-sl}

In this section, we extend the previous results to Strategy
Logic~(\SL). This logic has been initially introduced in~\cite{CHP07b}
for two-player turn-based games. It has then been extended to
$n$-players concurrent games in~\cite{MMV10a}. As~explained in the
introduction, satisfiability has been shown undecidable when
considering infinite structures~\cite{MMV10a}, and the proof
in~\cite{TW12} for finite satisfiability of \ATLsc straightforwardly extends to~\SL.
Here we show that satisfiability is decidable when considering
turn-based games and when fixing a finite alphabet, and that it
remains undecidable when only considering memoryless strategies.


\paragraph{Strategy Logic in a nutshell.}
%
We start by briefly recalling the main ingredients of \SL. The syntax is given by the following grammar: 
\[
\phi,\psi ::= p \mid \phi \et \psi \mid  \non \phi \mid  \X \phi \mid  
  \phi \Until \psi \mid \Diam[x] \phi \mid  (a,x)\phi
\]
where $a\in \Agt$ is an agent and $x$ is a (strategy) variable (we~use
$\Var$ to denote the set of these variables).
\looseness=-1
Formula $\Diam[x]\phi$ expresses the existence of a strategy, which is
stored in variable~$x$, under which formula~$\phi$ holds.  In~$\phi$,
the \emph{agent binding} operator $(a,x)$ can be used to bind
agent~$a$ to follow strategy~$x$.
An~assignment~$\chi$ is a partial function from $\Agt \cup \Var$ to~$\Strat$. 
 \SL~formulas are interpreted over pairs~$(\chi,q)$ where $q$ is a
 state of some \CGS and $\chi$~is an assignment such that every free
 strategy variable\slash agent\footnote{We use the standard notion of
   freedom for the strategy variables with the hypothesis that
   $\Diam[x]$ binds~$x$, and for the agents with the hypothesis that $(a,x)$
   binds~$a$ and that every agent in~$\Agt$ is free in
   temporal subformula (\ie, with $\Until$ or $\X$ as root).}
 occurring in the formula belongs to $\dom(\chi)$.
  Note that we have $\Agt \subseteq \dom(\chi)$ when temporal
  modalities $\X$ and~$\Until$  are interpreted: this implies
  that the set of outcomes is restricted to a unique execution
  generated by all the strategies assigned to players in $\Agt$, and
  the temporal modalities are therefore interpreted over this
  execution. Here we just give the semantics of the main two
  constructs (see~\cite{MMV10a} for a complete definition of~\SL):
%
\begin{xalignat*}1
\calC, \chi, q \sat \Diam[x] \phi &\quad \mbox{iff}\quad \exists F \in
\Strat \mbox{ s.t. } \calC,\chi[x \mapsto F], q \sat \phi \\ 
\calC, \chi, q \sat (a,x) \phi & \quad\mbox{iff} \quad
\calC,\chi[a\mapsto \chi(x)],q \sat \phi
\end{xalignat*}
In the following we assume w.l.o.g. that every quantifier~$\Diam[x]$
introduces a fresh strategy variable~$x$: this allows us to
permanently use variable~$x$ to denote the selected strategy for~$a$.

\paragraph{Turn-based case.}
%
\looseness=-1
The approach we used for \ATLsc can be adapted for~\SL. Given an \SL
formula $\Phi$ and a mapping $V\colon \Agt \to \Var$, we~define a
\QCTL* formula $\widehat{\Phi}^V$ inductively as follows (Boolean
cases omitted):
\begin{xalignat*}2
\widehat{\Diam[x] \phi}^V &= \exists \movsf_x. \Big[ \AG \Big( \EX_1 \movsf_x\Big) 
   \et  \widehat{\phi}^V\Big] &
\widehat{(a,x) \phi}^V &= \widehat{\phi}^{V[a\rightarrow x]}
\end{xalignat*}

Note that in this case we require that \emph{every} reachable state
has a (unique) successor labeled with~$\movsf_x$: indeed when one
quantifies over a strategy~$x$, the~agent(s) who will use this
strategy are not known~yet. However, in the turn-based case, a~given
strategy should be dedicated to a single agent: there is no natural
way to share a strategy for two different agents (or~the other way
around, any two strategies for two different agents can be seen as a
single strategy), as they are not playing in the same states. When the
strategy~$x$ is assigned to some agent~$a$, only the choices made in
the $a$-states are considered.

The temporal modalities are treated as follows:
\begin{xalignat*}1
\widehat{\phi\Until\psi}^V &= \All \Big[ \G \Big(\ET_{a_j\in \Agt}  
  (\turnsf_j \thn \X \movsf_{V(a_j)})\Big) \thn \widehat{\phi}^V \Until \widehat{\psi}^V \Big]
\\
\widehat{\X \phi}^V &= \All \Big[ \G   \Big(\ET_{a_j\in \Agt} 
  (\turnsf_j \thn \X \movsf_{V(a_j)})\Big) \thn \X \widehat{\phi}^V  \Big]
\end{xalignat*}
Now let $\widetilde{\Phi}$ be the formula $\Phi_{tb} \et \widehat{\Phi}^{V_\emptyset}$.
Then we have the following theorem:
%
\begin{restatable}{theorem}{thmsattbsl}
\label{thm-sat-tb-sl}
Let $\Phi$ be an \SL formula and $\widetilde{\Phi}$ be the \QCTL*
formula defined as above.  Then $\Phi$ is satisfiable in a turn-based
\CGS if, and only if, $\widetilde{\Phi}$~is satisfiable (in~the tree
semantics).
\end{restatable}

\paragraph{Bounded action alphabet}

Let $\Alac$ be $\{1,\ldots,\alpha\}$.  The reduction carried out for
\ATLsc can also be adapted for~\SL in this case.  Given an \SL
formula~$\Phi$ and a partial function $V\colon\Agt \rightarrow \Var$,
we define the \QCTL* formula $\widetriangle{\Phi}^V$ inductively as
follows:
\begin{xalignat*}2
\widetriangle{\Diam[x] \phi}^V &= \exists
\choosesf^1_x\ldots\exists \choosesf^\alpha_x. \AG \Big( \OU_{1\leq m
  \leq \alpha} \choosesf^m_x \et \ET_{n\not= m} \non \choosesf^n_x
\Big) \et  \widetriangle{\phi}^V\quad
&
\widetriangle{(a,x) \phi}^V &= \widetriangle{\phi}^{V[a\mapsto x]}
\end{xalignat*}

The temporal modalities are handled as follows:
\begin{xalignat*}1
\widetriangle{\phi\Until\psi}^V &= \All \Bigl[ \Bigl(\G \ET_{a_j\in
    \Agt} \ET_{1\leq m \leq \alpha} \bigl(\choosesf^m_{V(a_j)} \thn \X
  \movsf^m_{j}\bigr) \Bigr) \thn \Bigl( \widetriangle{\phi}^V \Until
  \widetriangle{\psi}^V \Bigr) \Bigr]
\\
\widetriangle{\X\phi}^C &= \All \Bigl[ \Bigl(\G \ET_{a_j\in \Agt} \ET_{1\leq m
    \leq \alpha} \bigl(\choosesf^m_{V(a_j)} \thn \X \movsf^m_{j}\bigr)
  \Bigr) \thn \Bigl( \X \widetriangle{\phi}^V \Bigr) \Bigr]
\end{xalignat*}
Remember that in this case, $\movsf^m_{j}$ labels the possible
successors of a state where agent $a_j$ plays $m$.  

Finally, let $\wideparen{\Phi}$ be the formula $\Phi_{\text{move}} \et
\widetriangle{\Phi}^V_\emptyset$. We~have:
\begin{restatable}{theorem}{thmsatbasl}
\label{thm-sat-ba-sl}
Let $\Phi$ be an \SL formula based on the set
$\Agt=\{a_1,\ldots,a_n\}$, let $\Alac=\{1,\ldots,\alpha\}$ be a finite
set of moves, and $\wideparen{\Phi}$ be the \QCTL* formula defined as above.
Then $\Phi$ is $(\Agt,\Alac)$-satisfiable if, and only if, $\wideparen{\Phi}$
is satisfiable (in the tree semantics).
\end{restatable}

\subsection{Memoryless strategies}
We now extend the undecidability result of \ATLsc0 to \SL with
memoryless-strategy quantification. Notice that there is an important
difference between \ATLsc0 and \SL0 (the logic obtained from \SL by
quantifying only on memoryless strategies):
the \ATLsc-quantifier $\Diamsz[A]$ still has an implicit quantification over
\emph{all} the strategies of the other players (unless their strategy is fixed
by the context), while in \SL0 all strategies must be explicitly quantified. 
Hence \SL0 and \ATLsc0 have uncomparable expressiveness.
Still:
\begin{theorem}
\SL0 satisfiability is undecidable, even when restricting to turn-based game
structures. 
\end{theorem}

\begin{proof}[Proof (sketch)]
The proof uses a similar reduction as for the proof for \ATLsc0. The
difference is that the implicitly-quantified strategies in \ATLsc0 are
now explicitly quantified, hence memoryless.  However, most of the
properties that our formulas impose are ``local'' conditions
(involving at most four nested ``next'' modalities) imposed in all the
reachable states. Such properties can be enforced even when
considering only the ultimately periodic paths that are outcomes of
memoryless strategies. The only subformula not of this shape is
formula $\All m \WUntil c$, but imposing this property along the
outcomes of memoryless strategies is sufficient to have the formula
hold true along any path.
\end{proof}

\section{Conclusion}

While satisfiability for \ATLsc and \SL is undecidable, we proved in
this paper that it becomes decidable when restricting the search to
turn-based games. We~also considered the case where strategy
quantification in those logics is restricted to memoryless strategies:
while this makes model checking easier, it~makes satisfiability
undecidable, even for turn-based structures. These results have been
obtained by following the tight and natural link between those
temporal logics for games and the logic \QCTL, which extends \CTL with
quantification over atomic propositions. This witnesses the power and
usefulness of \QCTL, which we will keep on studying to derive more results
about temporal logics for games.

\medskip

\paragraph*{Acknowledgement.} We thank the anonymous reviewers for their
numerous suggestions, which helped us improve the presentation of the paper.




\end{document}